\documentclass[aps,prb,reprint,superscriptaddress,longbibliography,floatfix]{revtex4-2}

\usepackage[T1]{fontenc}
\usepackage[utf8]{inputenc}
\usepackage{amssymb}
\usepackage{multirow}
\usepackage{graphicx}
\usepackage{amsmath}
\usepackage{color}
\usepackage{mathrsfs}
\usepackage{float}
\usepackage{indentfirst}
\usepackage{textcomp}
\usepackage{comment}
\usepackage{mathtools}
\usepackage{soul}
\usepackage[colorlinks=true,citecolor=blue,linkcolor=blue,urlcolor=blue]{hyperref}

\begin{document}
\title{The evolution of pairing correlation with $3d_{z^{2}}$ electron filling in a bilayer two-orbital model for La$_3$Ni$_2$O$_7$}

\author{Yufei Chen}
\affiliation{Key Laboratory of Artificial Structures and Quantum Control (Ministry of Education), School of Physics and Astronomy, Shanghai Jiao Tong University, Shanghai 200240, China}

\author{Yang Shen}
\affiliation{State Key Laboratory of Quantum Functional Materials, School of Physical Science and Technology, ShanghaiTech University, Shanghai 201210, China}

\author{Xiangjian Qian}
\affiliation{Key Laboratory of Artificial Structures and Quantum Control (Ministry of Education), School of Physics and Astronomy, Shanghai Jiao Tong University, Shanghai 200240, China}
\affiliation{Tsung-Dao Lee Institute, Shanghai Jiao Tong University, Shanghai 200240, China}

\author{Guang-Ming Zhang}
\email{zhanggm@shanghaitech.edu.cn}
\affiliation{State Key Laboratory of Quantum Functional Materials, School of Physical Science and Technology, ShanghaiTech University, Shanghai 201210, China}

\author{Mingpu Qin}
\email{qinmingpu@sjtu.edu.cn}
\affiliation{Key Laboratory of Artificial Structures and Quantum Control (Ministry of Education), School of Physics and Astronomy, Shanghai Jiao Tong University, Shanghai 200240, China}
\affiliation{Hefei National Laboratory, Hefei 230088, China}

\date{\today}

\begin{abstract}
The discovery of high-${T_c}$ superconductivity in pressurized bilayer nickelate La$_3$Ni$_2$O$_7$ presents a new arena for exploring unconventional pairing mechanisms. A pivotal yet unresolved issue is the specific role of the $3d_{z^{2}}$ orbital of Ni. While its inter-layer super-exchange antiferromagnetic coupling is widely considered crucial for superconductivity, the role of its itinerancy remains undetermined. Early studies showed that the superconductivity is accompanied by the emergence of a small Fermi pocket of the $3d_{z^{2}}$ orbitals. However, recent experiments show controversial results on the role of the $3d_{z^{2}}$ Fermi pocket on superconductivity. Motivated by these experimental results, we investigate an effective bilayer two-orbital model for La$_3$Ni$_2$O$_7$ using density-matrix renormalization group (DMRG) on a minimal one-dimensional geometry. By systematically varying the $3d_{z^{2}}$ orbital filling from $1/12$ doping to half-filling, we observe a pronounced suppression of superconducting correlations near half-filling. Our results demonstrate the itinerancy of $3d_{z^{2}}$ orbital is favorable for the pairing in the bilayer two-orbital model for La$_3$Ni$_2$O$_7$. Moreover, we observe that the pairing correlation is enhanced in regions where charge fluctuations are large, suggesting a competition between charge order and superconductivity in the model.
\end{abstract}

\maketitle

\section{Introduction} 

The recent discovery of high-${T_c}$ superconductivity in pressurized La$_3$Ni$_2$O$_7$~\cite{sun_signatures_2023} has attracted considerable attention in the Ruddlesden-Popper nickelates, following the discovery and subsequent development of superconductivity in infinite-layer nickelates \cite{li_superconductivity_2019,osada_nickelate_2021,zeng_superconductivity_2022}. La$_3$Ni$_2$O$_7$ provides a new platform and opportunity for investigating the microscopic mechanisms of high-${T_c}$ superconductivity, beyond the previous cuprate \cite{bednorz_possible_1986} and iron-based families \cite{kamihara_iron_based_2008}. Considerable experimental and theoretical progress has been made since this discovery, characterizing the properties of both its superconducting and normal states \cite{wang_normal_2024,wang_recent_2025}. More recently, superconductivity has also been observed in several La$_3$Ni$_2$O$_7$-based thin films at ambient pressure, including strained La$_3$Ni$_2$O$_7$ films, Pr-substituted (La,Pr)$_3$Ni$_2$O$_7$ films, and engineered hybrid films \cite{ko_signatures_2025,zhou_ambient_pressure_2025,nie_superconductivity_2026}.

La$_3$Ni$_2$O$_7$ is the $n=2$ member of the Ruddlesden--Popper nickelates La$_{n+1}$Ni$_n$O$_{3n+1}$, with NiO$_2$ bilayers separated by LaO layers. In each bilayer, two NiO$_6$ octahedra share an apical oxygen, forming an out-of-plane Ni--O--Ni structure, within which the Ni $3d_{z^2}$ orbital hybrids with the O $p_z$ orbital, resulting in an interlayer Ni--O--Ni $\sigma$-bond. First principles calculations show that pressure strengthens this interlayer hybridization and can drive a relatively flat $d_{z^2}$ bonding band across the Fermi level, generating a small $\gamma$ pocket \cite{luo_bilayer_2023,christiansson_correlated_2023,sakakibara_possible_2024}. In \cite{sun_signatures_2023,shen_effective_2023}, the emergence of superconductivity in La$_3$Ni$_2$O$_7$ was attributed to this metallization of the interlayer Ni--O--Ni $\sigma$-bonding band \cite{gao_finding_2015}.

Based on the calculated electronic structure, the low energy effective Hamiltonian for the pressurized La$_3$Ni$_2$O$_7$ can be modeled as a bilayer two-orbital model, involving the two Ni $e_g$ orbitals, $d_{x^2-y^2}$ and $d_{z^2}$ \cite{luo_bilayer_2023,shen_effective_2023,sakakibara_possible_2024}. In this model, the $d_{x^2-y^2}$ electron is constrained within the NiO$_2$ planes, while only inter-layer (Ni--O--Ni) hoppings are considered for the $d_{z^2}$ orbital given the special spatial characterization of these $d$ orbitals. Electronic structure calculation show that the $d_{z^2}$ orbital is close to half-filling while the $d_{x^2-y^2}$ orbital has a filling factor about $n=1/2$, resulting in a nominal Ni 3$d^{7.5}$ electronic configuration. This model and related ones have been widely studied recently \cite{gu_effective_2025,lechermann_electronic_2023,zhang_electronic_2023,yang_possible_2023,sakakibara_possible_2024,lu_interlayer_coupling_2024,yang_interlayer_2023,qu_bilayer_2024,shen_numerical_2025,chen_superconductivity_2025,fan_minimal_2025,oh_type_2023,jiang_high_temperature_2024,fan_superconductivity_2024,liu_s_wave_2023,qu_hunds_2025,yang_magnetism_2025}. Though the mechanism of superconductivity in La$_3$Ni$_2$O$_7$ is still under debate, most studies show the interlayer anti-ferromagnetic super-exchange of $d_{z^2}$ orbital is a key ingredient.

Recent discovery of superconductivity in La$_3$Ni$_2$O$_7$-based thin film at ambient pressure \cite{ko_signatures_2025,zhou_ambient_pressure_2025} enables ARPES measurements of the electronic structure in the superconducting phase, providing more insight to unveiling the mechanism of superconductivity in La$_3$Ni$_2$O$_7$ (for a review, see \cite{zhang_superconductivity_2026}). But in these ARPES measurements, whether the appearance of the $\gamma$ pocket is necessary for superconductivity is in controversy. Some experimental results support a positive connection between the $d_{z^2}$-derived $\gamma$ pocket and superconductivity: superconducting heterostructures show both $d_{x^2-y^2}$ and $d_{z^2}$ states are at the Fermi level, and engineered nickelate superstructures are superconducting when the $\gamma$ band crosses $E_F$ but not when it stays below $E_F$ \cite{li_arpes_2025,nie_superconductivity_2026}. Gap measurements on related superconducting films further show a nodeless superconducting gap on the observed Fermi surfaces \cite{shen_nodeless_2025}. Other ARPES experiments point to the opposite: superconducting La$_2$PrNi$_2$O$_7$ and Sr-doped La$_3$Ni$_2$O$_7$ films have been reported even though the $d_{z^2}$-derived $\gamma$ band remains about 70--75 meV below $E_F$ \cite{wang_electronic_structure_2025,sun_observation_2025}.

In this work, motivated by these controversial experimental results, we study the evolution of pairing correlation with the filling of $d_{z^2}$ orbital in an effective bilayer two-orbital model \cite{shen_effective_2023} for La$_3$Ni$_2$O$_7$. We tune the orbital chemical potential difference as a control parameter to redistribute carriers between the $d_{x^2-y^2}$ and $d_{z^2}$ orbitals with the total electron number fixed, thereby accessing regimes from finite $d_{z^2}$ hole density to near $d_{z^2}$ half filling. We perform large scale DMRG \cite{white_density_matrix_formulation_1992,white_density_matrix_algorithms_1993} calculations for a minimum one dimensional geometry of the model. Our accurate DMRG results show a pronounced suppression of superconducting correlations when the $d_{z^2}$ orbital is tuned to near half-filling. Our results demonstrate that itinerancy of the $d_{z^2}$ electrons favors superconductivity in the bilayer two-orbital model for La$_3$Ni$_2$O$_7$. We also find superconducting correlation is enhanced at regions where charge fluctuations are large, suggesting competition between charge order and superconductivity in the model.

The rest of this paper is organized as follows. Section~II introduces the effective bilayer two-orbital Hamiltonian and the lattice geometry we study. Section~III presents the DMRG results, including the evolution of orbital occupancy, charge and spin profiles, and orbital resolved pair-pair correlations. Section~IV discusses the missing term in our model. Section~V summarizes the implications for the role of itinerant $d_{z^2}$ electrons on pairing correlation in La$_3$Ni$_2$O$_7$. Additional numerical details and fitting procedures are provided in the Appendix.

\begin{figure}[htbp]
    \centering
    \includegraphics[width=\columnwidth]{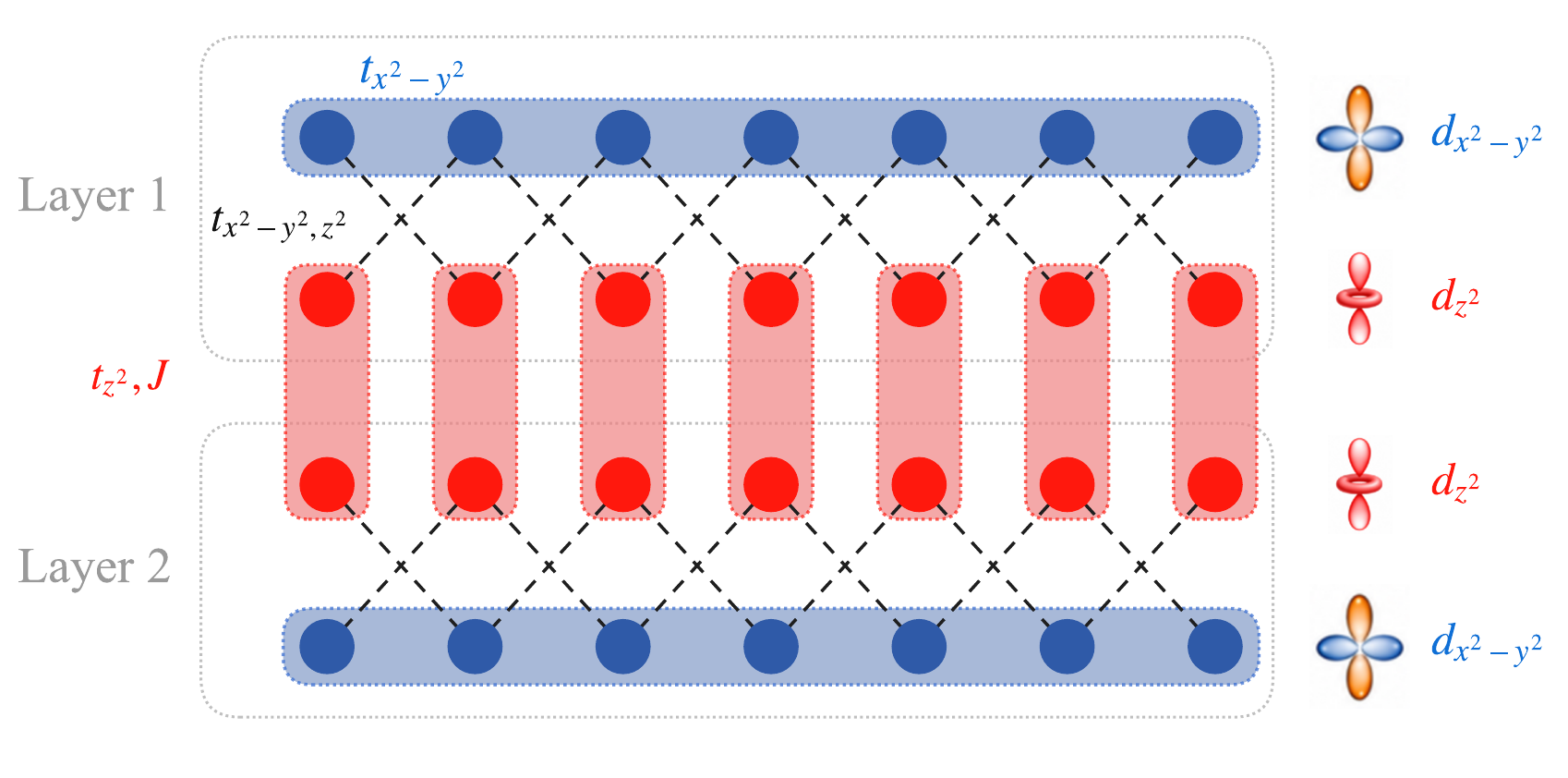}
    \caption{Schematic representation of the one-dimensional setup studied in this work for the bilayer two-orbital model. The blue sites denote the $d_{x^2-y^2}$ orbitals on the two outer legs, while the red sites denote the two $d_{z^2}$ orbitals in the central bilayer channel. The blue dashed rectangles indicate the nearest-neighbor hopping $t_{x^2-y^2}$ along the $d_{x^2-y^2}$ chains. The black dashed links represent the inter-orbital hybridization $t_{x^2-y^2,z^2}$, and the red dashed rectangles mark the inter-layer $d_{z^2}$ hopping and exchange channel, labeled $t_{z^2}$ and $J$.}
    \label{fig:lattice_schematic}
\end{figure}

\section{Model and Hamiltonian}

Based on the electronic structure, the low energy effective physics for pressurized La$_3$Ni$_2$O$_7$ can be modeled as a minimal bilayer two-orbital model. The two low-energy orbitals are the Ni $e_g$ orbitals: $d_{x^2-y^2}$ and $d_{z^2}$ orbitals \cite{luo_bilayer_2023,shen_effective_2023,shen_numerical_2025,yang_magnetism_2025,zhang_interlayer_2026,christiansson_correlated_2023,sakakibara_possible_2024}. The Hamiltonian takes the form
\begin{equation}
\begin{split}
    H = & -t_{x^2-y^2} \sum_{\langle ij \rangle, \sigma, a=1,2} (c^\dagger_{a,i,\sigma} c_{a,j,\sigma} + \text{h.c.}) \\
    & - \mu_{x^2-y^2} \sum_{i, a=1,2} n^c_{a,i} \\
    & - t_{x^2-y^2, z^2} \sum_{i, \sigma, a=1,2} (d^\dagger_{a,i,\sigma} \tilde{c}_{a,i,\sigma} + \text{h.c.}) \\
    & - t_{z^2} \sum_{i, \sigma} (d^\dagger_{1,i,\sigma} d_{2,i,\sigma} + \text{h.c.}) \\
    & + J \sum_i \boldsymbol{S}^d_{1,i} \cdot \boldsymbol{S}^d_{2,i} - \mu_{z^2} \sum_{i, a} n^d_{a,i},
\end{split}
\label{eq:model}
\end{equation}
where $c^\dagger_{a,i,\sigma}$ [$d^\dagger_{a,i,\sigma}$] creates an electron with spin $\sigma$ in the $d_{x^2-y^2}$ [$d_{z^2}$] orbital at site $i$ in layer $a$, and
$n^c_{a,i}=\sum_\sigma c^\dagger_{a,i,\sigma}c_{a,i,\sigma}$ and
$n^d_{a,i}=\sum_\sigma d^\dagger_{a,i,\sigma}d_{a,i,\sigma}$ are the corresponding density operators. We impose the constraint that double occupancy of the $d_{z^2}$ orbital is forbidden to account for the near half-filling of $d_{z^2}$ and large on-site Coulomb repulsion. The first line of Eq.~\eqref{eq:model} describes nearest-neighbor intralayer hopping of the $d_{x^2-y^2}$ orbital. The second line contains the orbital-dependent chemical potential for the $d_{x^2-y^2}$ orbital. The third line represents the hybridization between the two $e_g$ orbitals, with
\begin{equation}
    \tilde{c}_{a,i,\sigma}=c_{a,i+\hat{x},\sigma}+c_{a,i-\hat{x},\sigma}-c_{a,i+\hat{y},\sigma}-c_{a,i-\hat{y},\sigma},
\end{equation}
which encodes the $d_{x^2-y^2}$ form factor of the interorbital coupling. The fourth line describes the principal interlayer hopping of the $d_{z^2}$ orbital, while the last line contains the interlayer Heisenberg super-exchange coupling $J$ (through the apical oxygen) between the two $d_{z^2}$ spins and the chemical potential $\mu_{z^2}$ for the $d_{z^2}$ orbital.

Following the first principle calculation \cite{luo_bilayer_2023}, we take $t_{z^2} = 1$ as the energy unit and set $t_{x^2-y^2} = 0.8$ and $t_{x^2-y^2,z^2} = 0.4$. $J$ is fixed at $J = 0.5$ to access a regime with substantial interlayer super-exchange coupling. We ignore the Hubbard interaction on the $d_{x^2-y^2}$ orbital because this orbital remains quite dilute throughout the parameter range considered. We also omit the tiny interlayer hopping of the $d_{x^2-y^2}$ orbital and the intralayer hopping of the $d_{z^2}$ orbital due to the spatial characterization of these $d$ orbitals. The nominal Ni 3$d^{7.5}$ electronic configuration results in an average filling of $3/4$ with $d_{z^2}$ orbital close to half filling ($n=1$) and $d_{x^2-y^2}$ orbital close to a filling of $1/2$. In our numerical study, we scan the chemical potential difference
\begin{equation}
    \mu \equiv \mu_{x^2-y^2}-\mu_{z^2}
\end{equation}
to tune the actual filling of $d_{z^2}$ orbital, while the total number of electron is fixed to ensure a nominal Ni 3$d^{7.5}$ electronic configuration. Larger $\mu$ progressively shifts the $d_{z^2}$ orbital toward half filling at which the $d_{z^2}$ electron becomes insulating, allowing an investigation of the effect of the itinerancy of $d_{z^2}$ electron on the pairing correlation in this bilayer two-orbital model for La$_3$Ni$_2$O$_7$.

In this work, we study the bilayer two-orbital model defined in Eq. (\ref{eq:model}) on a minimum one-dimensional geometry as illustrated in Fig.~\ref{fig:lattice_schematic}. We perform large scale DMRG calculations on this setup with length $L_y=32$ with open boundary conditions.

\section{Numerical Results}

In our DMRG calculations, we retain bond dimensions up to $D=18000$, corresponding to a typical truncation error of $\epsilon \sim 10^{-6}$. Finite truncation error extrapolations are also performed to further improve the reliability of the DMRG results. For each parameter, we compute the orbital-resolved charge and spin responses and the pair-pair correlations. In this work, we mainly focus on the evolution of pair-pair correlations as the $d_{z^2}$ occupancy $\langle n_{d_{z^2}}\rangle$ is tuned from finite doping to near half-filling with $\langle n_{d_{z^2}}\rangle = 1$, to investigate the role of the itinerancy of $d_{z^2}$ electron on the superconductivity. 

\begin{figure}[h]
    \includegraphics[width=\columnwidth]{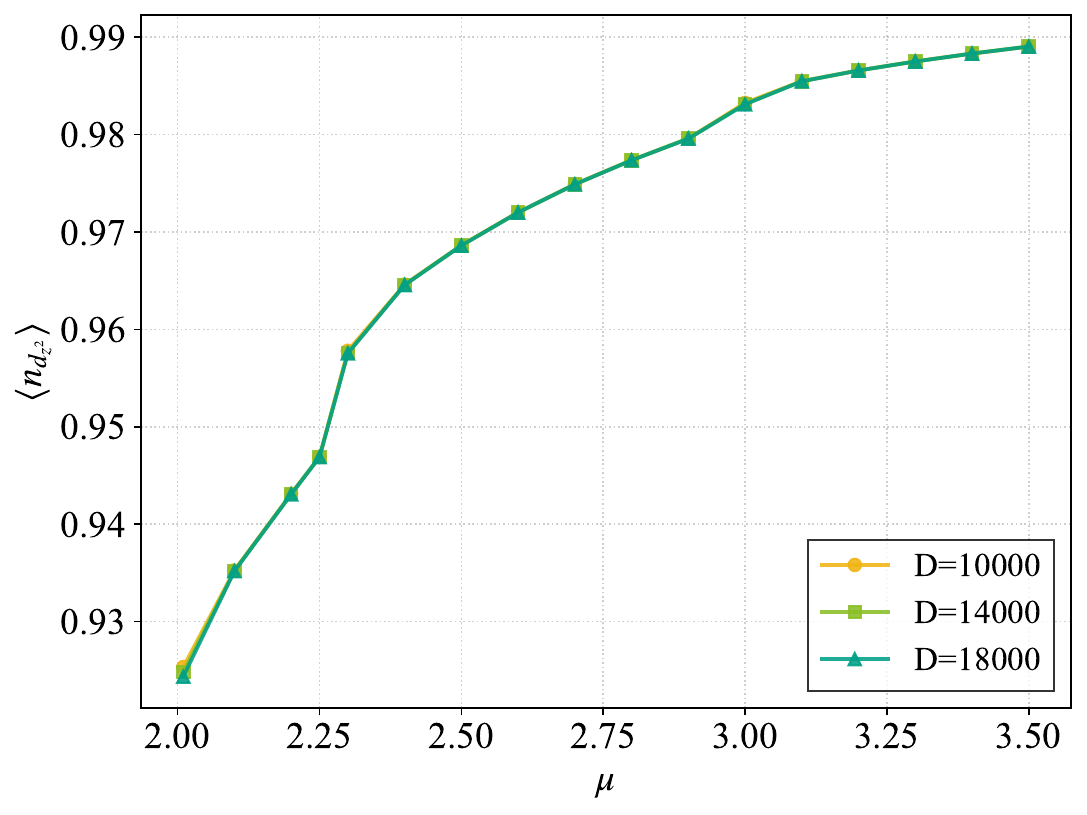}
    \caption{Orbital occupancy of the $d_{z^2}$ orbital as a function of the chemical potential difference $\mu$. Larger $\mu$ continuously shifts the $d_{z^2}$ orbital toward half filling ($\langle n_{d_{z^2}}\rangle=1$). The DMRG results with three bond dimensions nearly coincide, indicating that the orbital occupancy is well converged in the DMRG calculation. We notice that there exist two singularity points at $\mu \approx 2.3$ and $3.0$.}
    \label{fig:filling}
\end{figure}

We first show the mapping of $d_{z^2}$ occupancy $\langle n_{d_{z^2}}\rangle$ as a function of $\mu$ in Fig.~\ref{fig:filling}. As $\mu$ increases, the $d_{z^2}$ orbital evolves continuously from a moderately hole-doped regime with $h\approx1/12$ toward near half filling. Previous DMRG work \cite{shen_effective_2023} reported pronounced superconducting correlations at $\mu=2.01$, corresponding here to $\langle n_{d_{z^2}}\rangle=0.924$. We scan the $\mu$ up to $\mu=3.5$ (with corresponding $\langle n_{d_{z^2}}\rangle=0.989$) to track how the reduction of $d_{z^2}$ hole density is reflected in the pairing, charge, and spin responses. We notice that there exist two singularity points at $\mu \approx 2.3$ and $3.0$ in Fig.~\ref{fig:filling}. At these two points, the density has a dramatic change with a small change of $\mu$, indicating large charge fluctuation which could enhance pairing correlations, as we will discuss later.    

\begin{figure*}[htbp]
    \includegraphics[width=\textwidth]{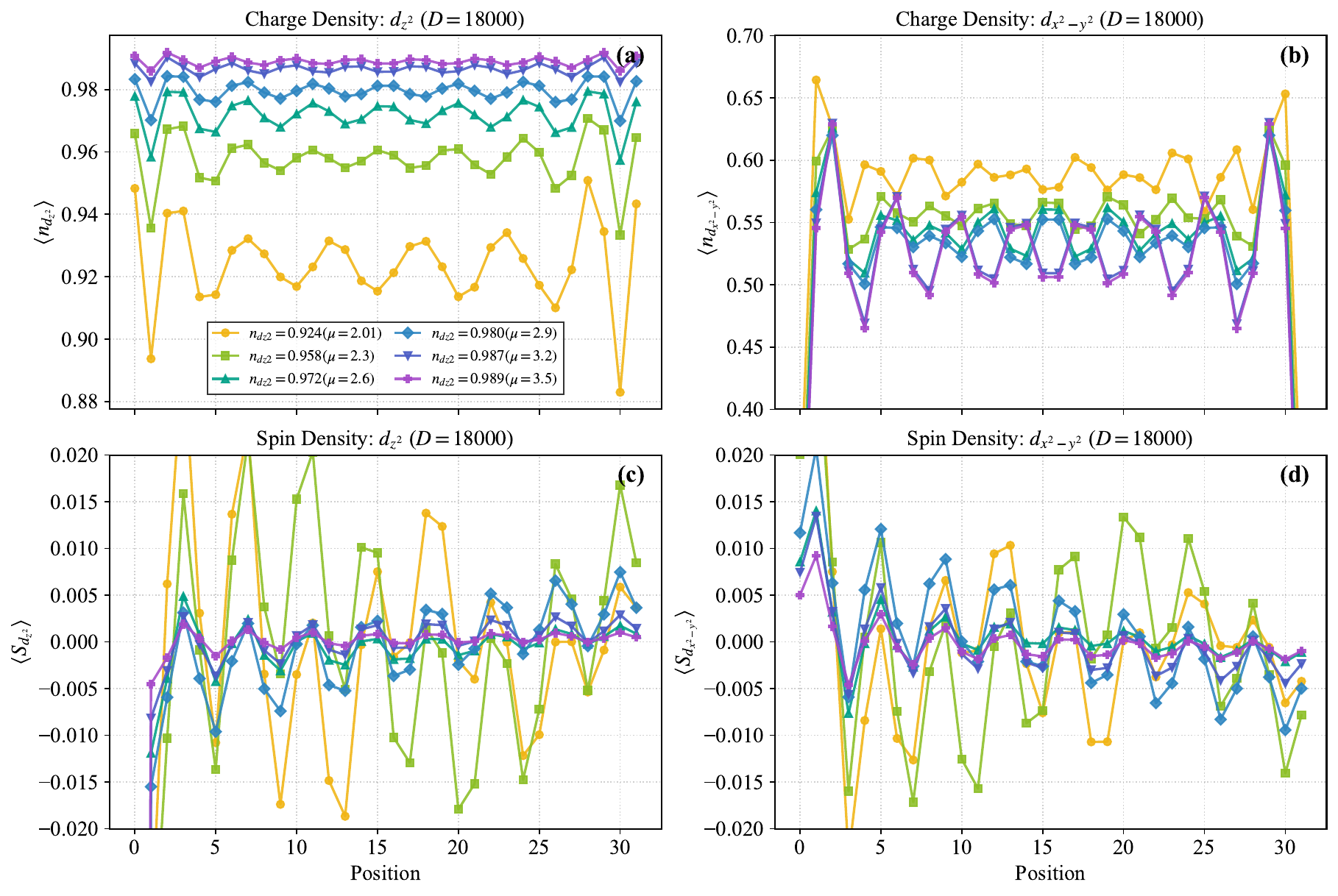}
    \caption{Charge and spin distributions of the $d_{z^2}$ orbital [(a),(c)] and the $d_{x^2-y^2}$ orbital [(b),(d)] for representative values of $\langle n_{d_{z^2}}\rangle$; the corresponding $\mu$ values are also shown in the legends. The DMRG results are with bond dimension up to $D=18000$ with truncation error $\epsilon\sim10^{-6}$ and convergence with bond dimension is checked. Charge density waves with wave-length $4$ and $5$ are observed in both orbitals. Around the two singularity points in Fig.~\ref{fig:filling} with $\mu \approx 2.3$ and $3.0$, the charge density wave pattern changes. See the main text for more discussion. Similar density wave pattern is observed in the spin channel but with weak amplitude.}
    \label{fig:charge_spin_dist}
\end{figure*}

\begin{figure*}[htbp]
    \includegraphics[width=\textwidth]{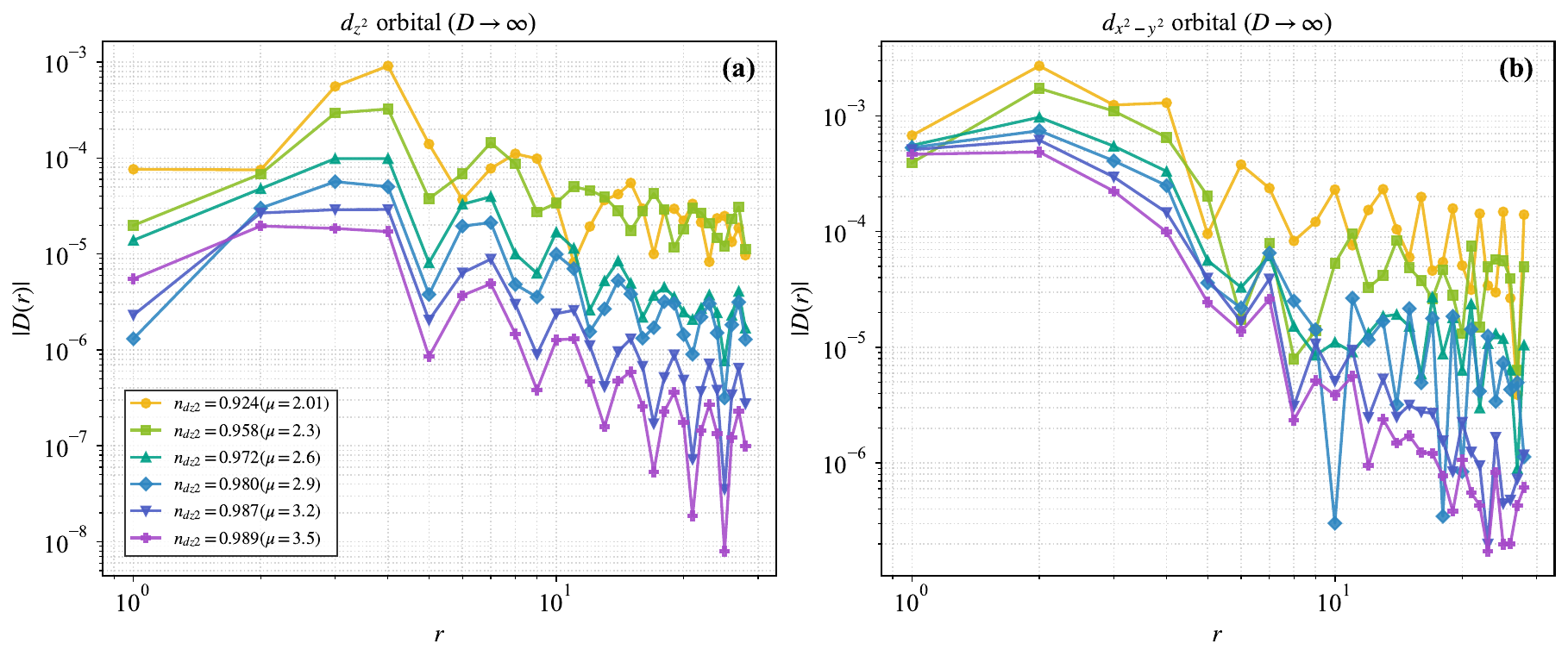}
    \caption{Pairing correlation functions of the $d_{z^2}$ and $d_{x^2-y^2}$ orbitals for selected values of $\langle n_{d_{z^2}}\rangle$; the corresponding $\mu$ values are also shown in the legends. The values are obtained by extrapolating finite-bond-dimension data to the zero-truncation-error limit $\epsilon \to 0$. In both orbitals, the correlations exhibit power-law decay as a function of distance, while the long-distance pairing amplitude is progressively suppressed as $\langle n_{d_{z^2}}\rangle$ increases toward half filling. Around the two singularity points in Fig.~\ref{fig:filling}, the pairing correlation is enhanced, consistent with the large charge fluctuation observed around these two points in Fig.~\ref{fig:charge_spin_dist}.}
    \label{fig:SC_dist}
\end{figure*}

\begin{figure}[htbp]
    \includegraphics[width=\columnwidth]{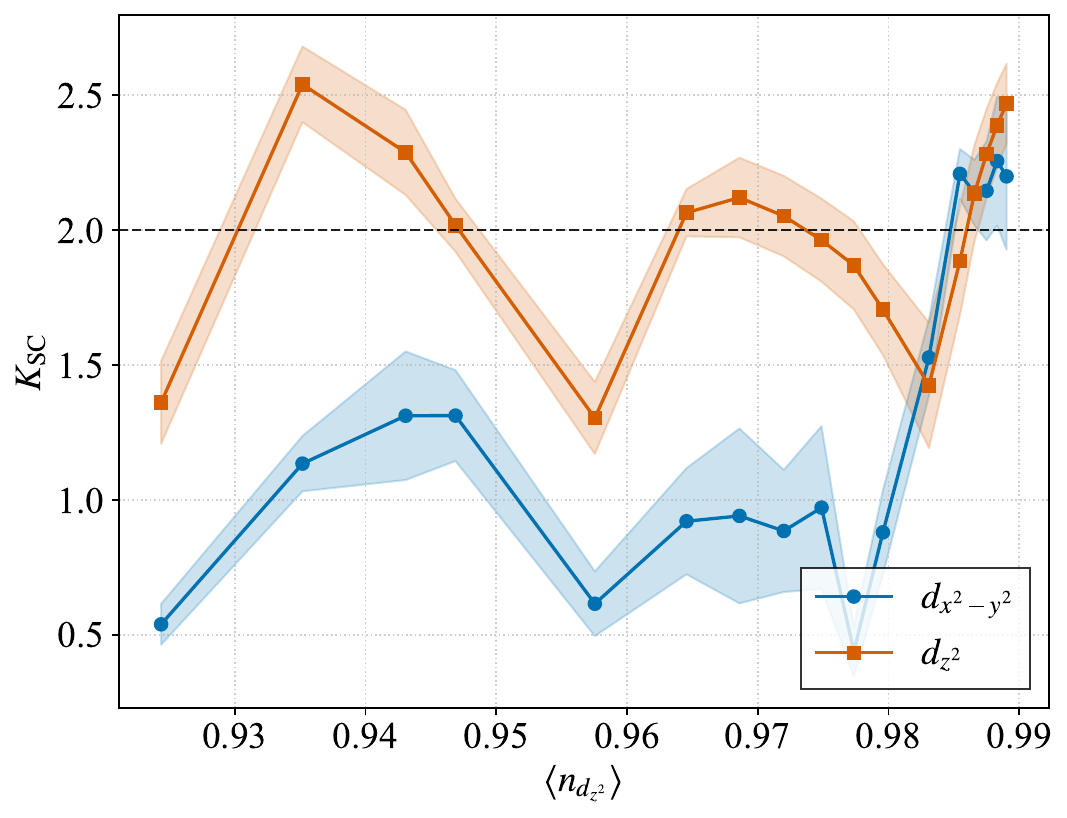}
    \caption{Power-law exponents $K_{SC}$ extracted from the pairing correlations of the $d_{z^2}$ and $d_{x^2-y^2}$ orbitals as functions of $\langle n_{d_{z^2}}\rangle$. The fits are performed using data extrapolated to the zero-truncation-error limit. Overall, $K_{SC}$ increases as the density in $d_{z^2}$ orbital is increased to half-filling for both the $d_{z^2}$ and $d_{x^2-y^2}$ orbitals with dips found in the two singularity points in Fig.~\ref{fig:filling}. A substantial increase of $K_{SC}$ occurs near $\langle n_{d_{z^2}}\rangle\approx0.983$ in both channels, indicating suppression of the superconducting tendency as the $d_{z^2}$ orbital approaches half filling. The shaded regions indicate fitting uncertainties. The dashed line marks $K_{SC}=2$, the quasi-one-dimensional criterion below which the pairing susceptibility would diverge for an asymptotic correlation function $D(r)\sim r^{-K_{SC}}$ in a Luther--Emery-like regime.}
    \label{fig:K_SC_summary}
\end{figure}

To probe the magnetic response, we apply a pinning magnetic field of amplitude $h_m=0.5$ at the left edge and analyze the induced spin response. Fig.~\ref{fig:charge_spin_dist} shows the charge and spin distributions of the $d_{z^2}$ and $d_{x^2-y^2}$ orbitals for representative values of $\langle n_{d_{z^2}}\rangle$ and corresponding $\mu$. For both the $d_{z^2}$ and $d_{x^2-y^2}$ orbitals, we can find charge density wave patterns with wavelength $4$ to $5$. Overall, with the increase of filling of $d_{z^2}$ orbital, the strength of the charge density waves is enhanced (suppressed) in the $d_{x^2-y^2}$ ($d_{z^2}$) orbital. Moreover, the two singularity points in Fig.~\ref{fig:filling} mark the positions where the charge density wave pattern is changed. Around the first singularity point with $\mu = 2.3 $ and $\langle n_{d_{z^2}}\rangle\approx 0.958$, the wave-length of charge density wave switches from $\lambda \approx 5$ to $\lambda \approx 4$, which can be clearly seen in the density profile of $\langle n_{d_{z^2}}\rangle$. Around the other singularity point with $\mu = 3.0 $ and $\langle n_{d_{z^2}}\rangle\approx 0.983$, the wave-length does not change, but the spatial phase for the charge density wave has a $\pi$ shift, which is clearly shown in the density profile of $\langle n_{d_{z^2}}\rangle$ and $\langle n_{d_{x^2-y^2}}\rangle$. More results on charge and spin density around these two singularities are shown in the Appendix. Similar density wave patterns are observed in the spin channel but with weak amplitude, especially in $\langle n_{d_{z^2}}\rangle \approx 1$ region.  

To characterize the superconducting tendency, we calculate the equal-time spin-singlet pairing correlation function between bond $i$, connecting sites $(i, 1)$ and $(i, 2)$, and bond $j$, connecting sites $(j, 1)$ and $(j, 2)$. The correlation function is defined as $D(i, j) = \langle \hat{\Delta}_i^\dagger \hat{\Delta}_j \rangle$, where the pairing operator is
\begin{equation}
    \hat{\Delta}_i^\dagger = \frac{1}{\sqrt{2}} (\hat{c}_{(i, 1), \uparrow}^\dagger \hat{c}_{(i, 2), \downarrow}^\dagger - \hat{c}_{(i, 1), \downarrow}^\dagger \hat{c}_{(i, 2), \uparrow}^\dagger).
\end{equation}
In evaluating these pair-pair correlations, we choose the vertical (inter-layer) $d_{z^2}$ bond at site $i=4$ as the reference bond. Fig.~\ref{fig:SC_dist} presents the pairing correlation functions of the inter-layer $d_{z^2}$ bonds and the intra-layer $d_{x^2-y^2}$ bonds for selected values of $\langle n_{d_{z^2}}\rangle$. DMRG results extrapolated to zero truncation error are shown. Throughout the parameter range considered, the correlations can be well described by a power-law form, $|D(r)|\sim r^{-K_{SC}}$, where $r=|i-j|$ denotes the distance between the two bonds. Overall, the long-distance pairing correlations are progressively weakened as $\langle n_{d_{z^2}}\rangle$ increases toward half filling. The pairing correlation in $d_{x^2-y^2}$ channel is stronger than that in the $d_{z^2}$ channel, consistent with the picture that the inter-layer super-exchange anti-ferromagnetic coupling of $d_{z^2}$ orbital provides pairing clue but the coherence of pairing is resulted from the very itinerant $d_{x^2-y^2}$ orbital \cite{shen_effective_2023}. Around the two singularity points in Fig.~\ref{fig:filling}, the pairing correlation is enhanced, consistent with the large charge fluctuation observed around these two points in Fig.~\ref{fig:charge_spin_dist}. As discussed above, these singularity points also mark the places where the charge order pattern is changed. So these results also suggest a competition between charge order and superconductivity in the model. More pairing correlation results around the two singularity points are shown in the Appendix.   

To quantitatively study the decay behavior of the pairing correlation, we perform a power-law fit of the results and summarize the fitted pairing exponents $K_{SC}$ for the $d_{z^2}$ and $d_{x^2-y^2}$ orbitals as functions of $\langle n_{d_{z^2}}\rangle$ in Fig.~\ref{fig:K_SC_summary}. Details of the fits can be found in the Appendix. Overall, $K_{SC}$ increases as the density in $d_{z^2}$ orbital is increased to half-filling for both the $d_{z^2}$ and $d_{x^2-y^2}$ orbitals with dips found at the two singularity points in Fig.~\ref{fig:filling}, consistent with the above analysis. In the vicinity very close to the $\langle n_{d_{z^2}}\rangle = 1$, exponents $K_{SC}$ for both $d_{z^2}$ and $d_{x^2-y^2}$ orbitals are larger than $2$, indicating the absence of superconductivity in this region. These results indicate that the itinerancy of $d_{z^2}$ orbital is favorable for the superconductivity in the bilayer two-orbital models for La$_3$Ni$_2$O$_7$. 

\section{Discussion}
In our model Hamiltonian defined in Eq.~(\ref{eq:model}), we treat the local Hilbert space of $d_{z^2}$ orbital as a t-J model like by imposing the constraint that double occupancy is forbidden. But in real material where the Coulomb repulsion is finite, when the $d_{z^2}$ is half-filled and charge degree of freedom is frozen, there should be an emergent Kondo-like coupling between the intra-layer $d_{z^2}$ and $d_{x^2-y^2}$ orbitals \cite{khaliullin_orbital_2026}. It will be interesting to also consider this Kondo coupling in our model in the future.

\section{Conclusion and Perspective}

In summary, we have investigated the evolution of pairing correlation with $d_{z^2}$ electron filling in an effective bilayer two-orbital model for La$_3$Ni$_2$O$_7$. By tuning the orbital chemical potential difference, we accessed regimes from moderate hole-doping of the $d_{z^2}$ orbital to near half-filling. For all the parameters we studied, we find that the pairing correlation in the $d_{x^2-y^2}$ orbital are stronger than that in the $d_{z^2}$ orbital. Our DMRG results also demonstrate a significant suppression of pairing correlations when the $d_{z^2}$ orbital approaches half-filling. This supports the picture proposed in \cite{shen_effective_2023}, where inter-layer antiferromagnetic super-exchange in the $d_{z^2}$ orbital provides the pairing glue, while hybridization with the more itinerant $d_{x^2-y^2}$ orbital transfers this pairing to establish long-range coherence. Our results indicate that the itinerancy of the $d_{z^2}$ orbital is favorable for superconductivity in the bilayer two-orbital model of La$_3$Ni$_2$O$_7$. Moreover, we find that the pairing correlation is enhanced in regions where charge fluctuations are large, suggesting a competition between charge order and superconductivity in the model.

\begin{acknowledgments}
We acknowledge the support from the National Natural Science Foundation of China (Grant No. 12522406 and No. 12274290), the Innovation Program for Quantum Science and Technology (2021ZD0301902), and the National Key Research and Development Program of MOST of China (No.2022YFA1405400 and No. 2023YFA1406400).
\end{acknowledgments}


\bibliography{merged_references_clean}
\clearpage
\onecolumngrid

\appendix*

\section{Additional numerical results and fitting procedures}
\label{app:details}

In the appendix, we provide additional numerical details including results for more chemical potential differences $\mu$ in the vicinity of the two singularity points in Fig.~\ref{fig:filling}, and the procedure used to extract the pairing exponent $K_{SC}$ from the pairing correlation functions.

\subsection{More results in the vicinity of the two singularity points in Fig.~\ref{fig:filling}}
\label{app:windows}
\begin{figure*}[htbp]
    \includegraphics[width=\textwidth]{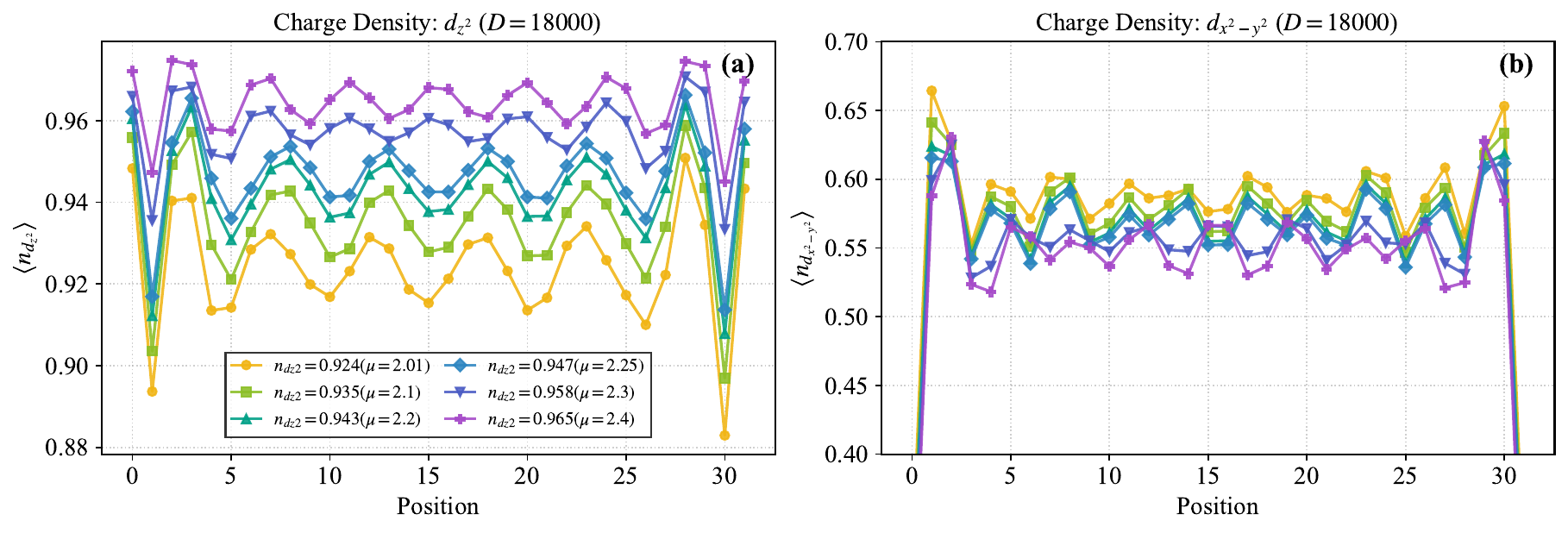}
    \caption{Charge distributions for $\langle n_{d_{z^2}}\rangle=0.924$--$0.965$ (corresponding to $\mu=2.01$--$2.4$). Panels (a) and (b) show the charge density of the $d_{z^2}$ and $d_{x^2-y^2}$ orbitals, respectively. We can clearly see a change of the charge density wave pattern at the singularity point around $\mu = 2.3$ in Fig.~\ref{fig:filling}.}
    \label{fig:charge_dist_region1}
\end{figure*}

\begin{figure*}[htbp]
    \includegraphics[width=\textwidth]{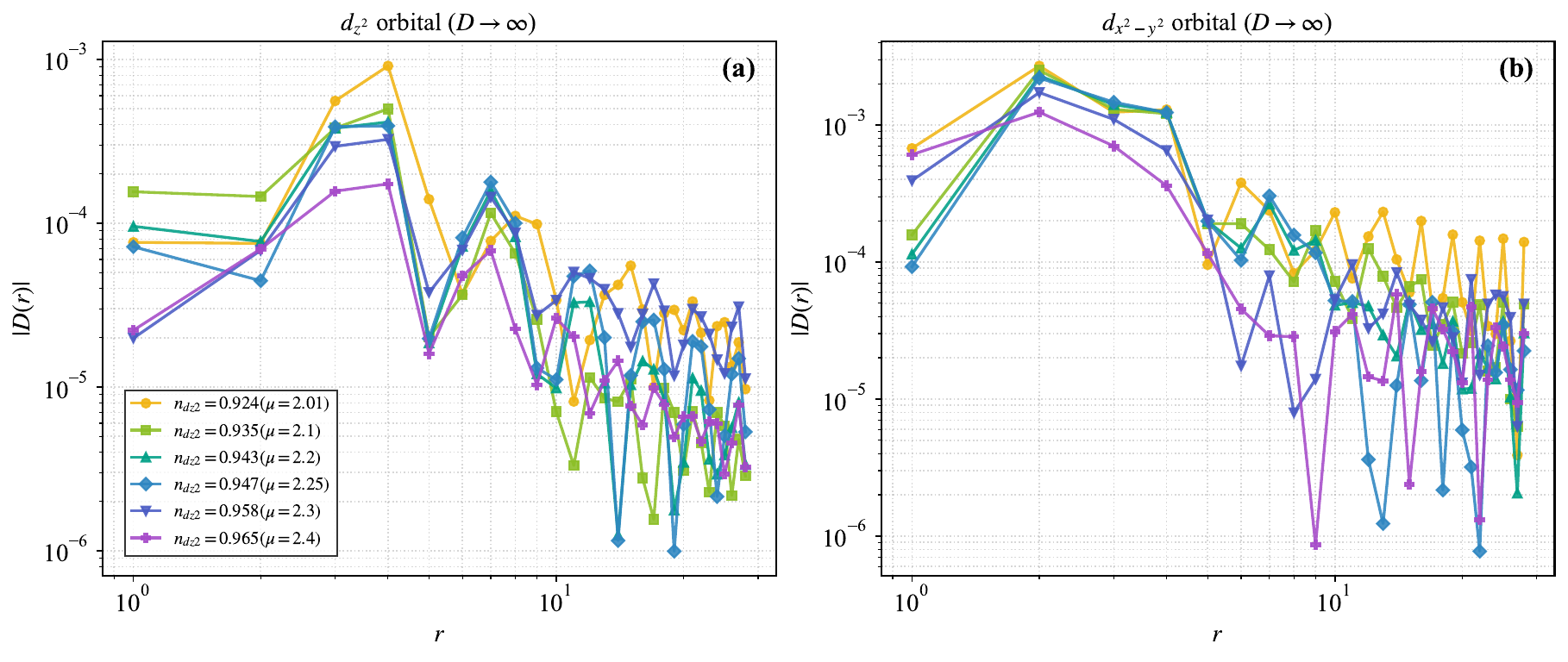}
    \caption{Pairing correlation functions for $\langle n_{d_{z^2}}\rangle=0.924$--$0.965$ (corresponding to $\mu=2.01$--$2.4$), for the $d_{z^2}$ orbital (a) and the $d_{x^2-y^2}$ orbital (b). We can clearly see an enhancement of pairing correlation at the singularity point around $\mu = 2.3$ in Fig.~\ref{fig:filling}. }
    \label{fig:SC_dist_region1}
\end{figure*}

\begin{figure*}[htbp]
    \includegraphics[width=\textwidth]{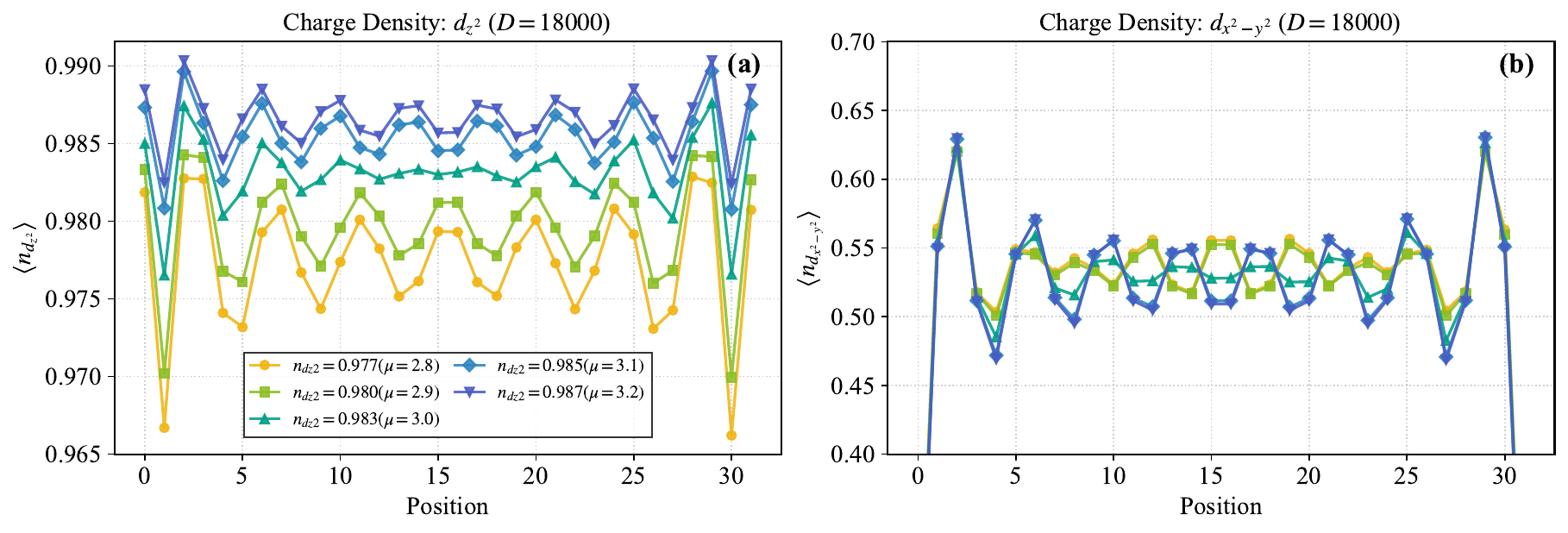}
    \caption{ Charge distributions for $\langle n_{d_{z^2}}\rangle=0.977$--$0.987$ (corresponding to $\mu=2.8$--$3.2$). Panels (a) and (b) show the charge density of the $d_{z^2}$ and $d_{x^2-y^2}$ orbitals, respectively. We can clearly see a change of the charge density wave pattern at the singularity point around $\mu = 3.0$ in Fig.~\ref{fig:filling}.}
    \label{fig:charge_dist_region2}
\end{figure*}

\begin{figure*}[htbp]
    \includegraphics[width=\textwidth]{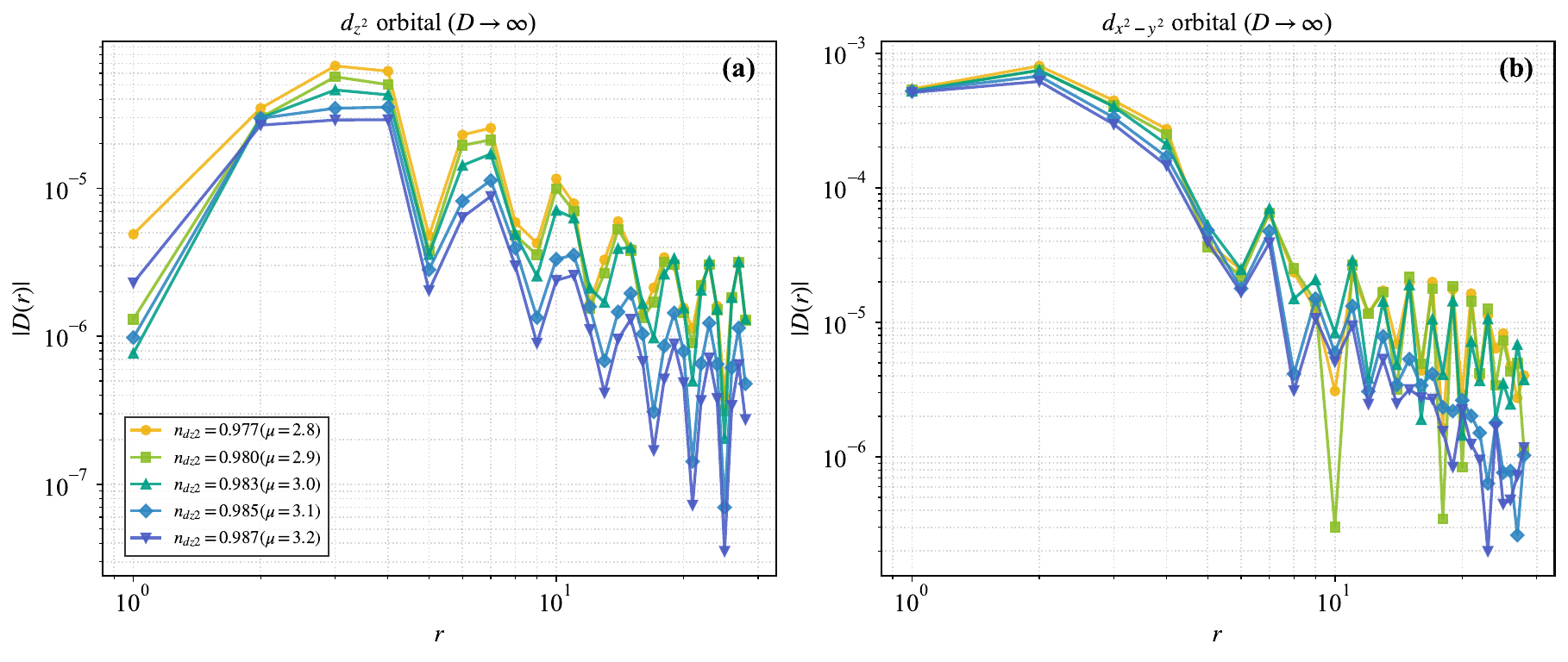}
    \caption{Pairing correlation functions for $\langle n_{d_{z^2}}\rangle=0.977$--$0.987$ (corresponding to $\mu=2.8$--$3.2$), for the $d_{z^2}$ orbital (a) and the $d_{x^2-y^2}$ orbital (b). We can clearly see an enhancement of pairing correlation at the singularity point around $\mu = 3.0$ in Fig.~\ref{fig:filling}.}
    \label{fig:SC_dist_region2}
\end{figure*}

In Figs.~\ref{fig:charge_dist_region1} and \ref{fig:SC_dist_region1}, we show more results in the vicinity of the singularity point at $\mu=2.3$. In Fig.~\ref{fig:charge_dist_region1}, we can clearly find the wavelength of the charge density wave changes from $\lambda \approx 5$ to $\lambda \approx 4$. Moreover, there is a spatial $\pi$ phase shift in the charge density wave pattern.  The large charge fluctuations in this singularity cause an enhancement of pairing correlation as shown in \ref{fig:SC_dist_region1}. Similar results at the other singularity point at $\mu=3.0$ are also shown in Figs.~\ref{fig:charge_dist_region2} and \ref{fig:SC_dist_region2}.

\subsection{Extrapolation to $D\to\infty$ and extraction of $K_{SC}$}
\label{app:fitting}

\begin{figure*}[htbp]
    \centering
    \includegraphics[width=\textwidth]{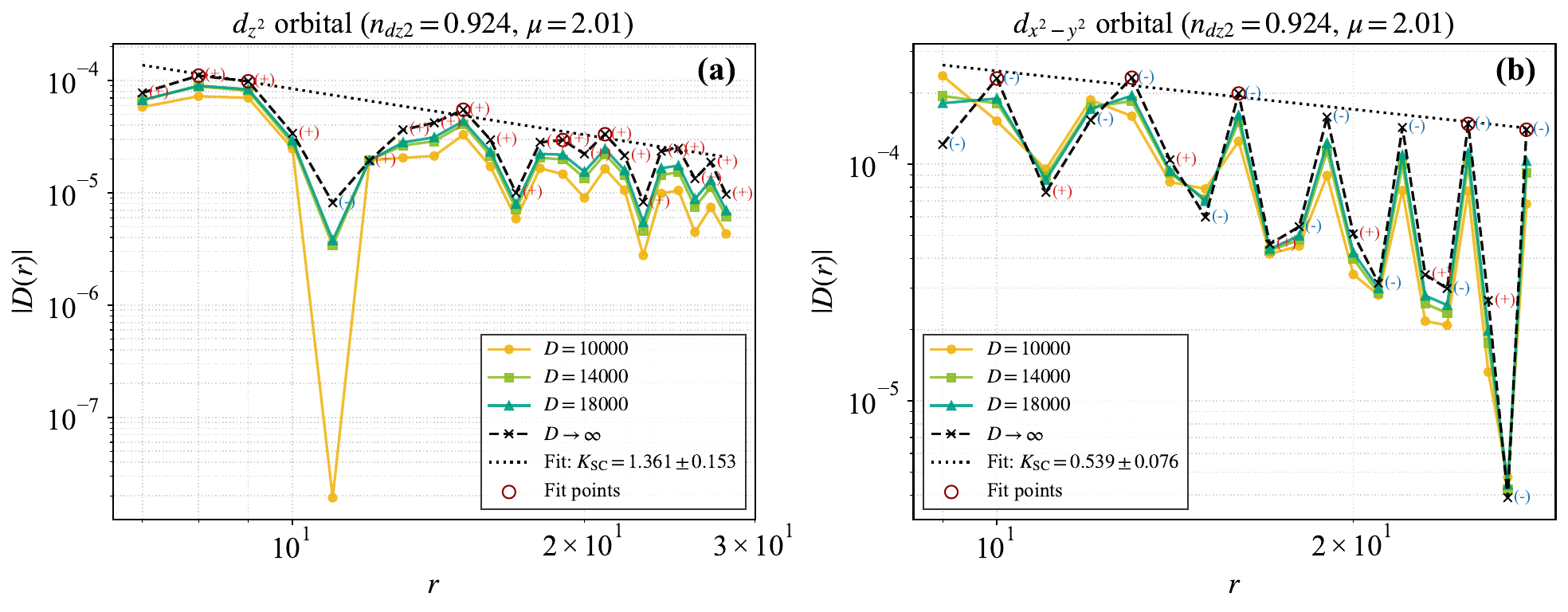}
    \caption{Representative extrapolation and fitting procedure for the pairing correlations at $\langle n_{d_{z^2}}\rangle=0.924$ ($\mu=2.01$). Panels (a) and (b) show the finite-$D$ data ($D=10000,14000,18000$), the extrapolated $D\to\infty$ results (black dashed curves), the fitted power-law envelopes (black dotted lines), and the fitting points (red circles) for the $d_{z^2}$ and $d_{x^2-y^2}$ orbitals, respectively. The $+$ and $-$ symbols mark the signs of the extrapolated pairing correlations before plotting $|D(r)|$. The $d_{z^2}$ correlations are almost all positive, whereas the $d_{x^2-y^2}$ correlations display a periodic sign-changing pattern, consistent with previous DMRG results \cite{shen_effective_2023}. The extracted exponents are $K_{SC}=1.361\pm0.153$ for the $d_{z^2}$ orbital and $K_{SC}=0.539\pm0.076$ for the $d_{x^2-y^2}$ orbital.}
    \label{fig:fit_mu201}
\end{figure*}

\begin{figure*}[htbp]
    \centering
    \includegraphics[width=\textwidth]{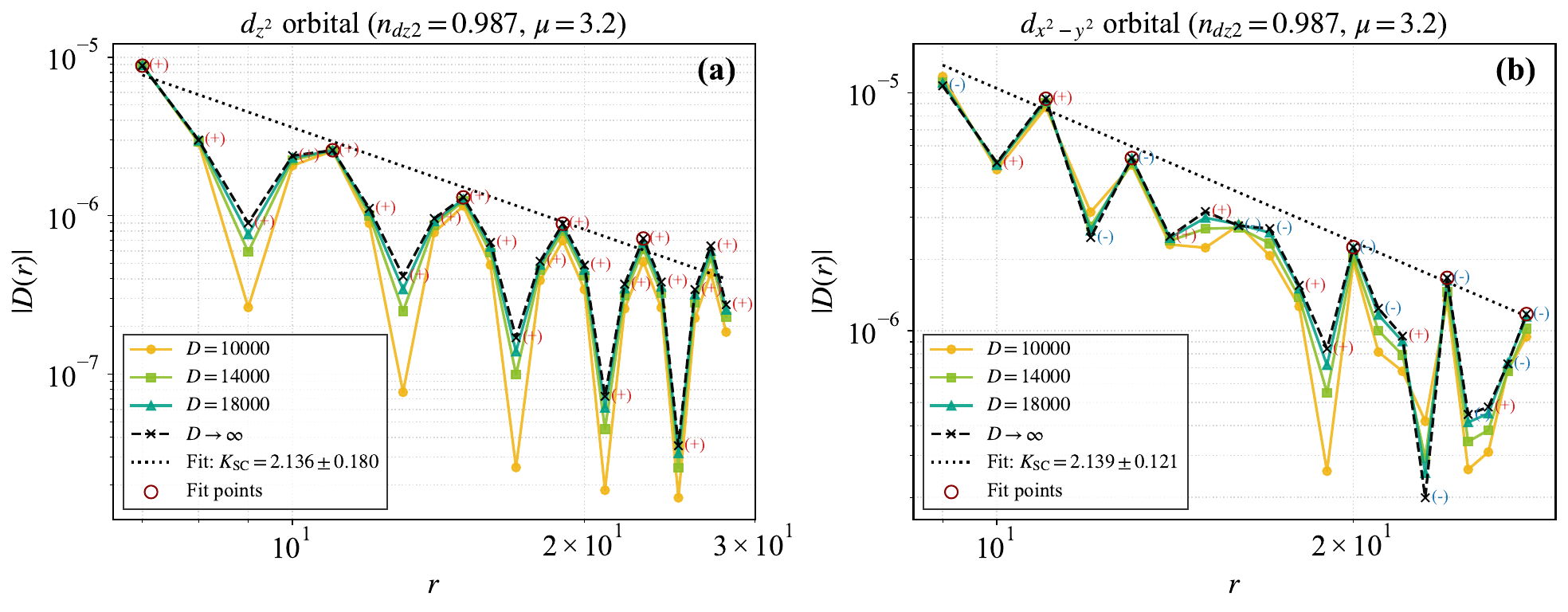}
    \caption{Representative extrapolation and fitting procedure for the pairing correlations at $\langle n_{d_{z^2}}\rangle=0.987$ ($\mu=3.2$). Panels (a) and (b) show the finite-$D$ data, the extrapolated $D\to\infty$ results, the fitted power-law envelopes, and the fitting points for the $d_{z^2}$ and $d_{x^2-y^2}$ orbitals, respectively. The $+$ and $-$ symbols mark the signs of the extrapolated pairing correlations before plotting $|D(r)|$. The $d_{z^2}$ correlations remain predominantly positive, whereas the $d_{x^2-y^2}$ correlations display a periodic sign-changing pattern. The extracted exponents are $K_{SC}=2.136\pm0.180$ for the $d_{z^2}$ orbital and $K_{SC}=2.139\pm0.121$ for the $d_{x^2-y^2}$ orbital. Compared with Fig.~\ref{fig:fit_mu201}, the larger exponents and reduced long-distance amplitudes indicate that superconducting correlations are suppressed in both orbitals when the $d_{z^2}$ orbital approaches half filling.}
    \label{fig:fit_mu32}
\end{figure*}

In DMRG calculations, we push the bond dimension to as large as $18000$ and the truncation error is at the order of $10^{-6}$. To further reduce the finite bond dimension error, we also perform finite truncation error extrapolation for the pairing correlation results as shown for two representative $\mu$ values ($\mu=2.01$ and $3.2$) in Figs.~\ref{fig:fit_mu201} and \ref{fig:fit_mu32}. 

The pairing exponent $K_{SC}$ is extracted by fitting the truncation error extrapolated DMRG results using a power-law form as $|D_{\infty}(r)| = A\, r^{-K_{SC}}$. Only the long-range pairing correlation results are used in the fit. In Figs.~\ref{fig:fit_mu201} and \ref{fig:fit_mu32}, panels (a) and (b) show the fitting procedure for the $d_{z^2}$ and $d_{x^2-y^2}$ pairing correlations for two representative $\mu$ values with $\mu=2.01$ and $3.2$. The data points used in the fit are indicated by red circles. The black dashed curves denote the extrapolated to zero truncation DMRG data, while the dotted black lines show the fitted power-law envelopes. The $+$ and $-$ symbols attached to the extrapolated data indicate the signs of the pairing correlations. The $d_{z^2}$ pairing correlations are predominantly positive, whereas the $d_{x^2-y^2}$ pairing correlations show a periodic sign-changing pattern. For $\langle n_{d_{z^2}}\rangle=0.924$ ($\mu=2.01$), this sign structure is consistent with previous DMRG results for the same model \cite{shen_effective_2023}.

\clearpage

\end{document}